# Ammonium adsorption, desorption and recovery by acid and alkaline treated zeolite

Sofia Maria Muscarella[1], Luigi Badalucco[1], Beatriz Cano[3], Vito Armando Laudicina[1*] and Giorgio Mannina[2]

[1] Department of Agricultural, Food and Forest Sciences, University of Palermo, Viale delle Scienze, Building 4, 90128 Palermo, Italy

[2] Department of Engineering, University of Palermo, Viale delle Scienze, Building 8, 90128 Palermo, Italy

[3] ZEOCEL ITALIA by DND Biotech srl, Via S. Cannizzaro 5, 56121 Pisa, Italia

*Corresponding Author: vitoarmando.laudicina@unipa.it; ph. +39 09123896556

**ABSTRACT**

In this study, the suitability of zeolite as a possible medium for ammonium adsorption, desorption and recovery from wastewater was investigated. Specifically, batch adsorption and desorption studies with solutions enriched in $NH_4^+$ were conducted employing zeolite to evaluate how the chemical treatment and contact time affect adsorption and desorption. Several experimental tests were carried out considering both untreated and treated zeolite. Untreated and HCl-Na treated zeolite adsorbed up to 11.8 mg $NH_4^+$ $g^{-1}$ and showed the highest efficiency in recovering $NH_4^+$ from aqueous solution. Regardless of pre-treatment, treatments with NaCl resulted in higher and faster adsorption of $NH_4^+$ than treatments with $CaCl_2$ and $MgCl_2$.

1. INTRODUCTION

Removal of nitrogen (N) from wastewater is relevant to limit eutrophication caused by the overuse of chemical fertilizers. In fact, the Water Framework Directive 2000/60/EC and the Council Directive 91/271/EEC have imposed strict consent limits (ammonium-N < 1 mg $L^{-1}$ and nitrate-N 10-30 mg $L^{-1}$ for discharge to fresh water and <50 mg $L^{-1}$ to seawater) to discharge N containing wastewater into the environment. Conventional removal techniques consist of biological process which allow to transform N into nitrite, nitrate and then nitrogen gases (Cruz et al., 2018). However, conventional biological removal processes because of the production of greenhouse gases may contribute to climate change (Mannina et al., 2017; 2018). To avoid such shortcomings $NH_4^+$ adsorption by adsorbents can be employed also due to its efficiency and overall simplicity compared to other methods (Canellas et al., 2019; Sengupta et la., 2015). Some advantages of the adsorption over the biological processes are i) shorter times of the adsorption process, ii) no reaction by-products, contaminants are directly removed from wastewater, iii) adsorbent can be regenerated easily and in-situ, iv) $NH_4^+$ enriched adsorbents can be used as slow-release fertilizer, and v) low operational costs (Sengupta., 2015). While adsorption is a promising method to tackle N removal from wastewater, finding suitable adsorbent materials which are abundant, low-cost and efficient remains a constant challenge (Han et al., 2021). Zeolite is the best candidate for N removal and can be a sustainable environmental vehicle for resource recovery from wastewater (Guaya et al., 2020; Han et al., 2021). Nevertheless, although zeolite is a low-cost material and can be use in-situ as filtering agent (Guida et al., 2021), some modification processes could be energy intensive and requires expensive chemical reagent, which may prohibit its large-scale application (Han et al., 2021). Zeolites are tectosilicates in which some $Si^{4+}$ ions are replaced by $Al^{3+}$ ions, thus resulting in an imbalance of net negative charges and improving adsorption ability (Moshoeshoe et al., 2017). The negative charges are counterbalanced by exchangeable cations lodge on the outer surface of the zeolite and held by weak electrostatic bonds (Widiastuti et al., 2011). Zeolites have a cation exchange capacity between 52 and 180 $cmol_{(+)}$ $kg^{-1}$. The wide variation range of the cation exchange capacity is due to mainly to type of zeolite affecting its structural and chemical properties, above all the $Si^{4+}/Al^{3+}$ ratio, and to particle diameter size (Sprynskyy et al., 2005a). To improve the adsorption capacity, natural zeolites may be chemically treated with acid, alkali, or salt (Wang and Peng, 2010) or can also be synthetized or engineered (Guida et al., 2020). The alkaline treatment aims to break all the covalent bonds between O and H of the OH groups in the tectosilicate structure, and to substitute $H^+$ with $Na^+$ thus creating an electronegative bond between O and Na. The acid treatment aims to leach all the exchangeable cations from the surface of the clinoptilolite replacing them with $H^+$ ions. Moreover, acid treatment should increase mesoporosity and microporosity by eliminating impurities even in the innermost cavities (Wang and Peng, 2010) thus increasing the adsorption ability of zeolite. The limitations of synthetized or engineered zeolites are the need of suitable equipment often not available in laboratories, long procedures, and costs (Han et al., 2021). Thus, there is a need to find less time-consuming and cheaper methods to improve the adsorption ability of zeolites. At the same time, zeolites should be also designed for the release of the adsorbed nutrients, such as $NH_4^+$ (Guaya et al., 2020). The aim of this study was to evaluate if the acid and alkaline treatments of zeolite can improve its ammonium adsorption and desorption ability. To our best knowledge, this is the first study assessing the effect of both acid and alkaline treatments on the same natural zeolite which is of paramount importance for N recovery (Wang and Peng, 2010). The final goal was to gain insights on the usage of zeolite for N recovery from wastewater in view of a better alignment to circular economy approach in the wastewater sector (Mannina et al., 2021).

## 2. MATERIALS AND METHODS

Natural Clinoptilolite (ZEOWATER ZN, Zeocel Italia by Dnd Biotech) was used. Clinoptilolite belongs to the heulandite group (HEU) whose tetrahedral structure consists of tetrahedral units of $SiO_4^{4-}$ and $AlO_4^{5-}$ and the chemical formula is $(Na,K)_6Al_6Si_{30}O_{72}$-$20H_2O$. It is constituted by 85% clinoplilotite, 8% cristobalite, 4% illite, 3-4% plagioclast and a Si/Al ratio of 4.8-5.5 and has bulk density of 0.98 g cm$^{-3}$, surface area of 40 m$^2$ g$^{-1}$, pH of 7.6. $K^+$ and $Ca^{2+}$ were the predominant exchangeable cations. Clinoptilolite was sieved to obtain a uniform diameter particle (ø 2.0 to 2.5 mm) and then washed three times with distilled water to remove particulate impurities on the surfaces and dried at 105°C for 2 hours. The chemical treatments applied to zeolite were 0.1M hydrochloric acid (HCl) or 1M sodium hydroxide (NaOH).

75 g of washed clinoptilolite were shaken three times with 200 mL of distilled water (UNT) or 1 M NaOH (NaOH) or 0.1M HCl (HCl) solution on an orbital shaker at 80 rpm for 4 h at room temperature, washed three times with 200 mL of distilled water to remove excess of NaOH or HCl. Finally, 25 g of treated clinoptilolite were shaken with 250 mL of distilled water or 1M NaCl, 0.5M $CaCl_2$ and 0.5M $MgCl_2$ for 24 h on an orbital shaker at 80 rpm at room temperature. After, samples were filtered through Whatman #42 filter paper, washed three times with 250 mL of distilled water to remove excess salts and dried at 105°C for 2 hours. The UNT samples were considered as control. To determine the $NH_4^+$ adsorption ability, 2 g of untreated and treated clinoptilolite were shaken with 200 mL of 1000 mg $NH_4^+$ $L^{-1}$ on orbital shaker for 24 h. After 24 h, samples were washed three times with 200 mL distilled water to remove excess $NH_4^+$ and placed in an oven for 2 h at 105 °C. From the initial 2 g, one gram of clinoptilolite was used to determine the amount of $NH_4^+$ adsorbed by Kjeldahl distillation with 30 mL of 33% NaOH solution for six minutes; the other gram was used to determine $NH_4^+$ desorption. In brief, 1 g of $NH_4^+$ enriched clinoptilolite was shaken with 100 mL of a 1M NaCl solution on a horizontal shaker for 48 hours at 80 rpm at room temperature. Then, it was washed three times with 200 mL of distilled water, oven-dried (2 h at 105°C) and analysed to determine retained $NH_4^+$ by Kjeldahl distillation (Gerhardt Vapodest 20 Distillation system, Netherlands). The amount of $NH_4^+$ desorbed was calculated as the difference between adsorbed (total) and retained (not exchanged by $Na^+$) $NH_4^+$. Ammonium adsorption kinetics by zeolites, during 48 h, were assessed by contacting 1 g of untreated and treated clinoptilolite with 100 mL of a 25 mg $NH_4^+$ $L^{-1}$ solution on a horizontal shaker at 80 rpm at room temperature. $NH_4^+$ concentration in aqueous phase was determined after 15, 30, 45 minutes, and 1, 2, 4, 8, 24, 48 hours by SEAL SFA AA100 Autoanalyzer (Alessandria, Italy). The percentage of $NH_4^+$ recovered from the solution compared to the $NH_4^+$ adsorbed by chemically treated zeolites was considered as the process efficiency (PE) and calculated as follows:

PE (%) = [(Adsorbed $NH_4^+$ – Retained $NH_4^+$) / $NH_4^+$ initial concentration]·100     (2)

Reported data were expressed on the oven dry (105°C) clinoptilolite weight basis. Data were subjected to one-way ANOVA with treatment as factor. Tukey test was carried out to assess significant differences at P<0.05 among treatments and control. Statistical analyses were performed using SPSS 13.0.

## 3. RESULTS AND DISCUSSION

### 3.1 Ammonium adsorption and desorption

Untreated and HCl-Na treated zeolite adsorbed the highest amount of $NH_4^+$, on average of 11.8 mg of $NH_4^+$ $g^{-1}$ (Fig. 1). The lowest amount of adsorbed $NH_4^+$ occurred with NaOH-Mg, that was 27% lower than untreated and HCl-Na treated clinoptilolite. The highest amount of $NH_4^+$ adsorbed by HCl-treated zeolite agreed with Sprynskyy et al. (2005b) who reported that mordenite treated with HCl and NaCl increased its ability in adsorbing $NH_4^+$. On the other hand, such results disagrees with Soetardji et al. (2015) who reported acid treatment of mordenite can cause dealumination, i.e. removal of $Al^{3+}$ ions from zeolite structure thus decreasing its ion exchange capacity. Actually, zeolite dealumination depends on pH of the solution. Hernández-Beltrán et al. (2008) clinoptilolite-rich tuff reported dealumination at pH 0, whereas at pH 4-6 changes in the crystallinity and dealumination were not observed. Considering that the pH of $NH_4Cl$ solution after 24 h of contact with HCl pre-treated clinoptilolite ranged from 3.5 to 3.8, we have hypothesized that no structural changes occurred to HCl-treated clinoptilolite thus maintaining the same adsorbing ability of the untreated one. The highest amount of adsorbed $NH_4^+$ was similar to that reported by Lebedynets et al., (2004). The latter, however, used clinoptilolite with a diameter range (0.16-0.315 mm) lower than that used in this study to adsorb $NH_4^+$ from a mono-component solution with an initial $NH_4^+$ concentration of 1000 mg $L^{-1}$. Considering that adsorption ability of zeolites increases by decreasing the particles diameter (Demir et al., 2002), such results suggest that clinoptilolite employed in this study has a greater ability in adsorbing $NH_4^+$ from mono-component solution. On the other hand, Lin et al. (2013) and Canellas et al. (2019)

using clinoptilolite with a lower diameter range (0.8-1.43 mm and 1.0-1.7 mm, respectively) found higher amount of $NH_4^+$ adsorbed, 17.1 and 20.7 mg $NH_4^+$ $g^{-1}$, respectively. The chemical treatments applied, except for HCl-Na treatment, decreased the adsorbed $NH_4^+$ per unit of mass of clinoptilolite. The reason why $NH_4^+$ adsorption was decreased by the chemical treatments may be ascribed mainly to two different causes. First, clinoptilolite treated with $CaCl_2$ and $MgCl_2$ following the treatment with HCl decreased its ability in adsorbing $NH_4^+$ probably due to a less efficient exchange occurring between the monovalent $NH_4^+$ and the divalent $Ca^{2+}$ or $Mg^{2+}$. This because the ability of a cation to displace another cation adsorbed on an exchanger decreases as the charge of the adsorbed cation increases, due to the increase in the electrostatic interaction force between the divalent cation and the negative surface charges on the exchanger (Essington, 2015). This highlights the importance of the exchangeable ions used in the treatment of zeolites and agree with simulations that have shown $Na^+$ producing lower energy state within the lattice than $Ca^{2+}$ (Canellas et al., 2019). Second, NaOH pre-treated zeolites had pH in water of 10.5, whereas the $NH_4Cl$ solution after 24 h of contact with NaOH pre-treated zeolite had pH ranging from 8.1 to 8.5. Being zeolite stable at alkaline pH and thus excluding any structural or chemical modification (Soetardji et al., 2015) the lower amount of adsorbed $NH_4^+$ may be due to the scarce ability of NaOH in by eliminating impurities in the innermost cavities of clinoptilolite or, to a lesser extent, to the volatilization of ammonia. Such a result agrees with Lebedynets et al. (2004) who reported that for optimum operations, the pH of the aqueous solution must be maintained at or below 7. The amount of $NH_4^+$ released after 48 h ranged from 76 to 83% of that previously adsorbed (Fig. 1) and depended on the amount of $NH_4^+$ adsorbed. Indeed, $NH_4^+$ adsorbed and retained were positively correlated ($R^2$=0.432; P<0.001). Those percentages are, on average, 10% lower than that reported by Canellas et al. (2019) that used clinoptilolite with a lower diameter range. Such different results can be due to greater affinity for $NH_4^+$ by tested clinoptilolite or to a greater adsorption of $NH_4^+$ within pores making the exchange with $Na^+$ more difficult. Overall, zeolite is a suitable, low cost and environmental friendly $NH_4^+$ adsorbent from wastewater and its use should be encouraged, although when used for multiple cycles to adsorb and desorb $NH_4^+$ its mechanical strength has to be evaluated (Guida et al., 2020).

### 3.2 Ammonium recovery from solution after desorption: process efficiency

Untreated and HCl-Na treated clinoptilolite showed the highest efficiency (9.4 % for both zeolites), whereas NaOH-pre-treated and NaOH-Mg the lowest (7.9 and 7.1%, respectively). The treatment NaOH-Ca had an efficiency higher than NaOH-Mg likely depended on the greater hydration sphere of $Mg^{2+}$ compared to $Ca^{2+}$. Such results are in line with those concerning the $NH_4^+$ adsorption and confirm that the alkaline pre-treatment did not improve the efficiency of clinoptilolite in recovering $NH_4^+$ from solution.

### 3.3 Ammonium adsorption kinetics

The amount of $NH_4^+$ adsorbed (Fig. 2) increased with contact time as reported in literature (Karadag et al., 2006). Within the first 15 minutes, UNT clinoptilolite adsorbed 12% of $NH_4^+$, HCl pre-treated clinoptilolite adsorbed from 8.6 to 11.2% of $NH_4^+$, whereas NaOH pre-treated from 0.2 to 11.6%. These results disagree with what found by Karadag et al. (2006) and Kotoulas et al. (2019) who reported more than 70% of $NH_4^+$ adsorption within the first 10 minutes using natural clinoptilolite. Both studies, however, used zeolites smaller in diameter (1.0-1.4 mm and 0.7-1.0 mm, respectively) than that used in this study (ø 2.0-2.5 mm). Thus, these different results can be ascribed to the different particle dimensions having the smaller one higher specific area and hence greater ability in $NH_4^+$ adsorption (Lin et al., 2013). More than 50% of $NH_4^+$ was adsorbed within 4 h (Fig. 2). After 4 h the adsorption rate started to slow down, probably because of changes in the mechanisms of adsorption: from diffusion on the outer surface to pore diffusion in the intra-particle matrix (Dimova et al., 1999). The maximum $NH_4^+$ adsorption was showed by HCl-Na clinoptilolite at 24 and 48 h. Kinetic adsorption of $NH_4^+$ followed different patterns depending on the pre-treatment. $NH_4^+$ adsorption by NaOH pre-treated and treated clinoptilolite, except for NaOH-Mg was almost completed after 8 hours, whereas that by HCl pre-treated clinoptilolite within 24 h for all treatments. These results suggested that NaOH-pre-treated clinoptilolite are generally faster in saturating their exchange sites with $NH_4^+$ than HCl-pre-treated treatments. Such rapid saturation can be explained by a greater presence of ionic bonds between O of zeolite and Na than covalent ones following the alkaline treatment. However, towards the end of the kinetic adsorption the amount of $NH_4^+$ adsorbed was on average lower with NaOH pre-treated clinoptilolite compared to untreated or HCl pre-treated one (Fig. 2). This was likely due to losses of $NH_4^+$ as ammonia as a consequence of high pH as previously reported. Nevertheless, these results are of greater importance to speed up the $NH_4^+$ adsorption process from solution although low pH has to be maintained to avoid $NH_3$ volatilization (Lin et al., 2013). Moreover, regardless of the pre-treatment, clinoptilolite showed different $NH_4^+$ adsorption kinetics depending on the salts used to saturate its exchange sites. The lower $NH_4^+$ adsorption rate on $Ca^{2+}$ and $Mg^{2+}$ treated clinoptilolite may be due to the stronger bonds that divalent cations establish with the exchanger because of their higher positive specific charge compared to monovalent cations and thus to their difficult displacement (Semmens and Martin, 1988).

## 4. CONCLUSIONS

Ammonium adsorption, desorption and recovery by clinoptilolite was studied in batch test. The alkaline treatment worsened the ability of clinoptilolite in adsorbing $NH_4^+$, whereas the acid treatment did not improve their ability in adsorbing $NH_4^+$. The amount of $NH_4^+$ desorbed and adsorbed was positively correlated and affected $NH_4^+$ recover. Clinoptilolite pre-treated with NaOH adsorbed less $NH_4^+$ and was saturated faster than that pre-treated with HCl. Treatments with NaCl resulted in higher and faster adsorption of $NH_4^+$ than treatments with $CaCl_2$ and $MgCl_2$.

E-supplementary data for this work can be found in e-version of this paper online.

**ACKNOWLEDGEMENTS**

This work was funded by the project "Achieving wider uptake of water-smart solutions—WIDER UPTAKE" (grant agreement number: 869283) financed by the European Union's Horizon 2020 Research and Innovation Programme, in which the last author of this paper, Giorgio Mannina, is the principal investigator for the University of Palermo. The Unipa project website can be found at: https://wideruptake.unipa.it/ Authors thank Dr. Giulio Giannardi and Zeocel Italia by Dnd Biotech for providing the natural clinoptilolite.

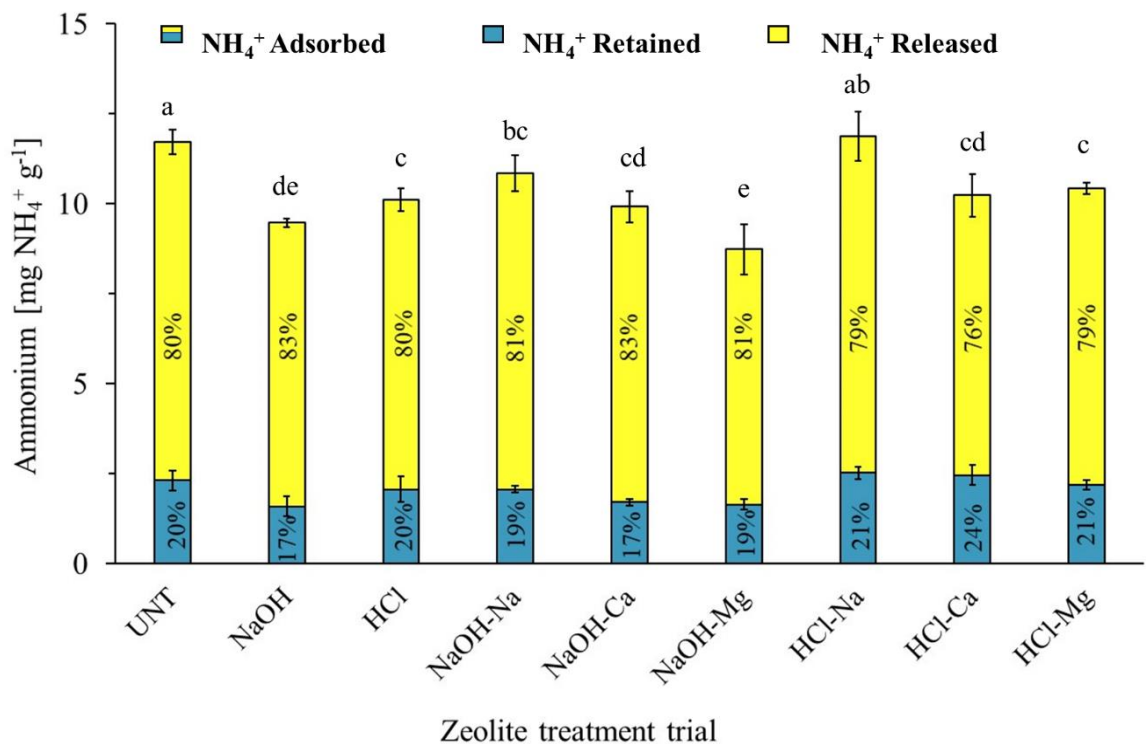

**Fig. 1.** Amount of ammonium adsorbed, and relative percentage of ammonium released and retained by treated clinoptilolite. Treatments are: untreated clinoptilolite (UNT), pre-treated clinoptilolite with 1M sodium hydroxide (NaOH) and 0.1M hydrochloric acid (HCl), and treated clinoptilolite with sodium chloride (NaCl), calcium chloride (CaCl$_2$) magnesium chloride (MgCl$_2$) after each pre-treatment. Values are mean of three replicates and bars are standard deviations. Different letters indicate significant differences at P<0.05 among treatments.

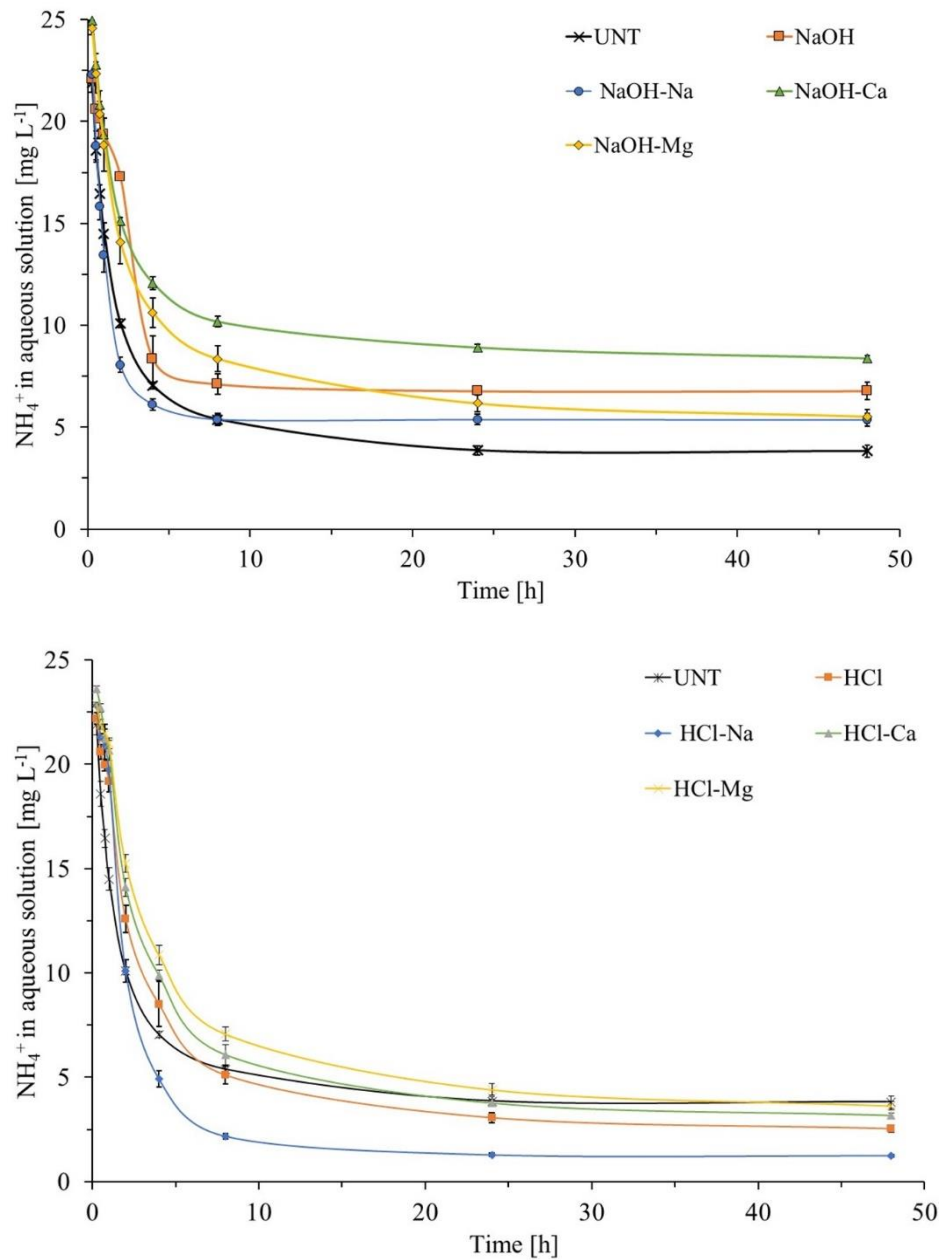

**Fig. 2.** Ammonium adsorption kinetics by treated clinoptilolite during 48 hours. Treatments are: untreated clinoptilolite (UNT), pre-treated clinoptilolite with 1M sodium hydroxide (NaOH) and 0.1M hydrochloric acid (HCl), and treated clinoptilolite with sodium chloride (NaCl), calcium chloride ($CaCl_2$) magnesium chloride ($MgCl_2$) after NaOH and HCl pre-treatment. Values are mean of three replicates and bars are standard deviations...